\newcommand{\rxy}{\rho_{xy}}
\newcommand{\rxx}{\rho_{xx}}

\newcommand{\sxy}{\sigma_{xy}}
\newcommand{\sxx}{\sigma_{xx}}

\documentclass[prb,twocolumn,showpacs,floatfix,amsmath,amssymb]{revtex4}

\usepackage{graphicx}

\begin{document}

\preprint{Ver.I}

\title{The effect of carrier density gradients on magnetotransport data
measured in Hall bar geometry}

\author{L.~A.~Ponomarenko}
\email{leonidp@science.uva.nl}
\author{D.~T.~N.~de~Lang}
\author{A.~de~Visser}

\affiliation{Van der Waals-Zeeman Institute, University of
Amsterdam, Valckenierstraat 65, 1018 XE Amsterdam, The
Netherlands}

\author{V.~A.~Kulbachinskii}
\affiliation{Low Temperature Physics Department, Moscow State
University, 119899, Moscow, Russia }

\author{G.~B.~Galiev}
\affiliation{Institute of Ultrahigh  Frequency Semiconductor
Electronics, RAS, 117105, Moscow, Russia}

\author{H.~K\"{u}nzel}
\affiliation{ Fraunhofer-Institut f\"{u}r Nachrichtentechnik,
Heinrich-Hertz-Institut, 10587 Berlin, Germany}

\author{A.~M.~M.~Pruisken}
\affiliation{Institute for Theoretical Physics, University of
Amsterdam, Valckenierstraat 65, 1018 XE Amsterdam, The
Netherlands}

\begin{abstract}
We have measured magnetotransport of the two-dimensional electron
gas in a Hall bar geometry in the presence of small carrier
density gradients. We find that the longitudinal resistances
measured at both sides of the Hall bar interchange by reversing
the polarity of the magnetic field. We offer a simple explanation
for this effect and discuss implications for extracting
conductivity flow diagrams of the integer quantum Hall effect.
\end{abstract}

\pacs{73.43.Qt, 73.43.Nq}

\date{\today}

\maketitle

\section{Introduction}

The primary technique for probing the two-dimensional electron gas
in semiconductor heterostructures is magnetotransport. In
practice, this is realized by passing a constant current $I$
through a Hall bar, i.e. a rectangular sample with several contact
pads for sensing the longitudinal voltage $V_{xx}$ and the
transverse or Hall voltage $V_{xy}$. Hall bars are routinely
prepared from semiconductor wafers by photo-lithographic
techniques. Typically, the    channel length is in the order of
2000~$\mu$m, with the distance between the longitudinal voltage
contacts $L~\sim$~500~$\mu$m and the channel width $W$ a factor
2-5 smaller. For a homogeneous sample three voltage contacts are
sufficient to determine the longitudinal and transverse
resistances, $R_{xx} = V_{xx}/I$ and $R_{xy}= V_{xy}/I$. However,
in the presence of macroscopic sample inhomogeneities, i.e.
inhomogeneities on a scale much larger than the typical
microscopic length scales of the electron gas, the resulting
resistances present average values. If the length scale of the
inhomogeneities is comparable to the sample size, information
about the inhomogeneities can be obtained by placing additional
contacts on the sample. An example of a macroscopic inhomogeneity
is a spatial variation in the carrier density, such as a small
gradient along the channel direction. Such gradients are not
uncommon in Hall bars and arise directly from the growth process.
Besides, the spatial variation might depend on the cooling
procedure.

\begin{figure}[b]
\includegraphics[width=7cm]{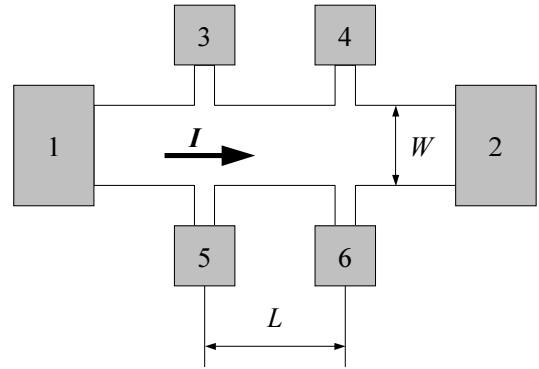}
\caption{\label{fig:scatch} Schematic picture of a Hall bar. $L$
denotes the distance between the longitudinal voltage contacts and
$W$ the channel width.}
\end{figure}
\begin{table*}
\caption{\label{table} Transport parameters of Hall bars, prepared
from different semiconductor structures, before and after
illumination: $L$ is the distance between longitudinal voltage
contacts, $W$ is the width of the Hall bar channel, $n_e$ is the
electron density, $\Delta n_e / n_e$ is the electron density
difference measured at different Hall contact pairs and $\mu$ is
the transport mobility. QW and HS stand for quantum well and
heterostructure, respectively.}
\begin{ruledtabular}
\begin{tabular}{cccccc}
System&Sample&$L\times W$, &$n_e$ &
$\Delta n_e / n_e$ & $\mu$ \\
 & &$(\mu m)^2$& (10$^{11}$ cm$^{-2}$)& &(cm$^2/$Vs) \\ \hline GaAs/AlGaAs QW&\#659 before illum.&$1260 \times 1000$&4.7&0.017&220 000\\
&\#659 after illum.&$1260 \times 1000$&6.1&0.0025&300 000
\\GaInAs/AlGaAs QW&31232-\#3 before illum.&$387 \times
75$&1.8&0.014&19 000\\ &31232-\#3 after illum.&$387 \times
75$&3.6&0.006&25 000\\ &31232-\#2&$387 \times 75$&2.2&0.018&25 000
\\GaInAs/InP HS&\#2&$1070 \times
650$&2.2&0.016&16 000
\end{tabular}
\end{ruledtabular}
\end{table*}
The influence of macroscopic sample inhomogeneities, geometrical
effects and contacts on magnetotransport data taken from Hall bars
has been investigated by a number of authors. For example,
von~Klitzing and Ebert \cite{vKlitzingGradient} investigated a
Hall bar with a fairly large carrier density variation
($\sim$10\%). They reported differences in the longitudinal
resistances measured on both sides of the Hall bar, as well as a
strong dependence of the amplitude of the Shubnikov-de Haas
oscillations on the magnetic field polarity. Geometrical effects
of Hall bars, such as the influence of the channel width, the
position of contacts, etc., have been investigated in detail by
Haug \cite{haug:sst93}, who observed, for instance, an asymmetric
Shubnikov-de Haas effect on a long Hall bar without noticeable
carrier density gradients. This effect persisted when the
direction of the magnetic field was reversed and was attributed to
the proximity of the voltage and current contacts on the Hall bar.
The influence of the contact geometry and size has been studied
experimentally by \textit{e.g.} Woltjer \textit{et al}.
\cite{woltjer:prb88} These authors explained a number of unusual
observations in terms of a local resistivity tensor.

In the course of low-temperature magnetotransport study of the
quantum Hall effect, we noted an interesting phenomenon in the
$R_{xx}$ data: upon reversing the polarity of the magnetic field
the longitudinal resistances measured on top and bottom sides of
the Hall bar, $R_{xx}^t$ and $R_{xx}^b$, interchange. Different
values of $R_{xx}^t$ and $R_{xx}^b$, are quite common for Hall
bars and are generally attributed to sample inhomogeneities
($R_{xx}^t = R_{xx}^b$ for a perfect Hall bar). However, the
interchange of $R_{xx}^t$ and $R_{xx}^b$ upon field reversal
cannot be accidental and requires a non-trivial explanation.
Moreover, the interchange of $R_{xx}^t$ and $R_{xx}^b$ was found
to be a generic feature of the samples investigated.

In this paper we present magnetotransport data obtained from
various quantum wells and heterostructures. All samples showed the
interchange of $R_{xx}^t$ and $R_{xx}^b$ upon field reversal
(except one heterostructure where contact misalignment was
dominant). We offer a simple explanation for this phenomenon,
namely a small carrier gradient along the channel direction of the
Hall bar. To the best of our knowledge, these specific
observations and the corresponding explanation, have not been
reported in literature before. Our results are important for the
study of conductivity flow diagrams of the quantum Hall effect.
They show that complications may arise when critical
conductivities are extracted in the standard fashion from
experimentally obtained resistivities on the plateau-plateau
transitions.

\section{Experimental}

Magnetotransport experiments have been performed on different
semiconductor structures: high-mobility ($\mu \sim$ 300000
cm$^2/$Vs) GaAs/AlGaAs quantum wells and low-mobility ($\mu \sim$
20000 cm$^2/$Vs) GaInAs/AlInAs quantum wells, as well as
InGaAs/InP heterostructures. The electron densities for all our
samples were fairly low ($n_e = 1.8 - 6.1 \times 10^{11}$
cm$^{-2}$), such that all samples showed distinct quantum Hall
features within our available magnetic field range ($B \leq$ 8 T).
Mobilities and electron densities of the samples are listed in
Table~\ref{table}. In some cases the electron density was
increased by illumination with a LED at $T =$ 4.2 K. Samples were
selected such that no carrier relaxation occurred during the
measurements. Throughout this paper we label the voltage contacts
on the Hall bar as sketched in Fig.~\ref{fig:scatch}, where
$V_{xx}^t$ refers to contacts 3-4 and $V_{xx}^b$ to 5-6.The
experiments were performed in an adsorption-pump operated $^3$He
cryostat, equipped with a superconducting magnet ($B_{max} =$ 8
T). The longitudinal and transverse resistance were measured
simultaneously, using standard lock-in techniques at a frequency
of 13 Hz. The excitation current ranged from 5 to 50 nA. The data
presented here were all taken at $T =$ 0.4 K.

\section{MAGNETOTRANSPORT RESULTS: ANTISYMMETRY IN $R_{xx}$}

\begin{figure}[htb]
\includegraphics[width=7.6cm]{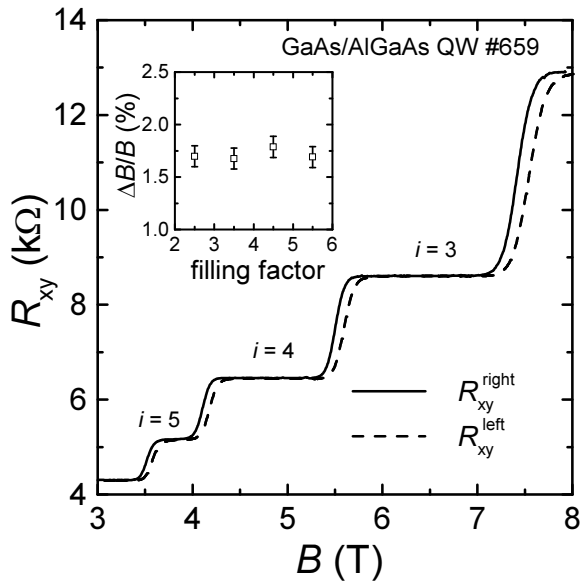}
\caption{\label{fig:Hall GaAs} Hall resistance as a function of
magnetic field for the GaAs/AlGaAs quantum well (\#659 - no
illumination) near plateaux $i = 2-6$ at $T = 0.4$ K. The solid
and dashed lines show data taken at the right and left Hall
contact pairs. The insert shows the relative difference between
the plateau-plateau transition fields $\Delta B / B$ as function
of the filling fraction.}
\end{figure}
In Fig.~\ref{fig:Hall GaAs} we show the Hall resistance between 3
and 8 T, before illumination, of the GaAs/AlGaAs quantum well
(sample \#659) measured at the left (3-5) and right (4-6) contact
pairs across the Hall bar. The left and right Hall resistance
traces, $R_{xy}^l = R_{xy}^{35} = (V_3 - V_5) / I$ and $R_{xy}^r =
R_{xy}^{46} = ( V_4 - V_6 ) / I$, were measured during the same
field sweep. Upon reversing the magnetic field, $R_{xy}^l$ and
$R_{xy}^r$ stay identical, except for the change of sign as it
should: $R_{xy}^{l,r} (B) = - R_{xy}^{l,r}(-B)$. This implies that
the contribution of $R_{xx}$ to the Hall resistance is absent and
that misalignment of Hall voltage contacts is negligible. The
plateau-plateau (PP) transitions measured at contacts 4-6 are
shifted along the field axis with respect to those at contacts
3-5. This is due to different local filling factors, $\nu (x,y) =
h e n_e(x,y)/B$. Assuming a constant magnetic field over the
sample, the spatial distribution of filling factors matches that
of the electron density $n_e$. Hence the filling factor between
contacts 3-5 is always larger than the one between 4-6. However,
their ratio is field independent and equal to the ratio of local
densities. This is corroborated by the insert in
Fig.~\ref{fig:Hall GaAs}, where we have traced the experimental
values of the shift $\Delta B / B$, measured halfway the PP
transitions, versus filling factor $2 < \nu < 6$. We extract a
carrier density difference $\Delta n_e / n_e = \Delta B / B =
0.017$. For simplicity, we will assume, in the remainder of the
paper, that the carrier gradient along the channel is constant.

\begin{figure}[htb]
\includegraphics[width=7.6cm]{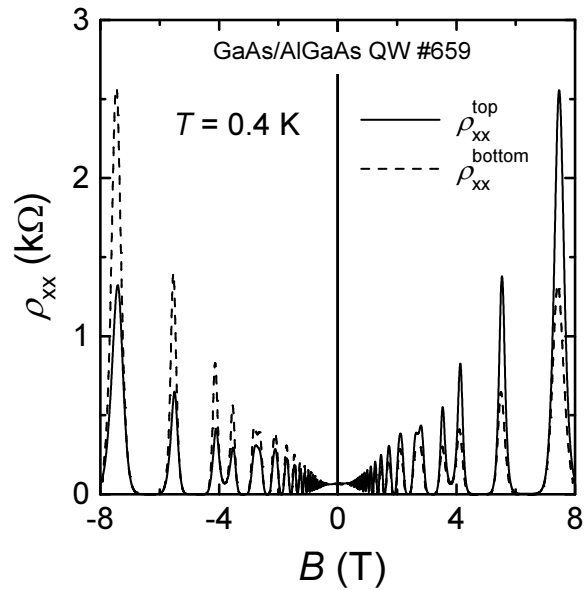}
\caption{\label{fig:Long GaAs before} Longitudinal resistivity as
a function of magnetic field ($B\uparrow$ and $B\downarrow$) for a
GaAs/AlGaAs quantum well (\#659 - no illumination). The solid and
dashed lines show data taken at contact pairs located at the top
and bottom side of the Hall bar. Upon reversing the magnetic field
we observe an antisymmetry in the longitudinal resistivity
components $\rxx^t(-B)= \rxx^b(B)$.}
\end{figure}

Next we present in Fig.~\ref{fig:Long GaAs before} the
longitudinal resistances $R_{xx}^t = R_{xx}^{34} = (V_3 - V_4) /
I$ and $R_{xx}^b = R_{xx}^{56} = (V_5-V_6) / I$ of sample \#659,
measured at top and bottom sides of the Hall bar.
Data are taken for field up ($B\uparrow$) and down ($B\downarrow$)
directions ($|B| \leq 8$ T). The data show the familiar
Shubnikov-de Haas oscillations at low fields and the resistance
peaks associated with the PP transitions in the quantum Hall
regime at higher fields. For a homogeneous Hall bar we expect
$R_{xx}^b = R_{xx}^t$, which is clearly not the case here. Instead
we find a large difference in the peak values $R_{xx}^b$ and
$R_{xx}^t$, which amounts up to 50\% for the $i =  3 \rightarrow
2$ transition. Moreover, a close inspection of Fig.~\ref{fig:Long
GaAs before} shows that $R_{xx}^t$ for $B\uparrow$ equals
$R_{xx}^b$ for $B\downarrow$ and \textit{vice versa} to within
1\%. We conclude that the longitudinal resistance when measured on
both sides of the Hall bar shows a remarkable
\textit{``antisymmetry''}:
\begin{equation}
\label{antisymmetry} R_{xx}^t (B) = R_{xx}^b (- B).
\end{equation}

\begin{figure}[htb]
\includegraphics[width=7.6cm]{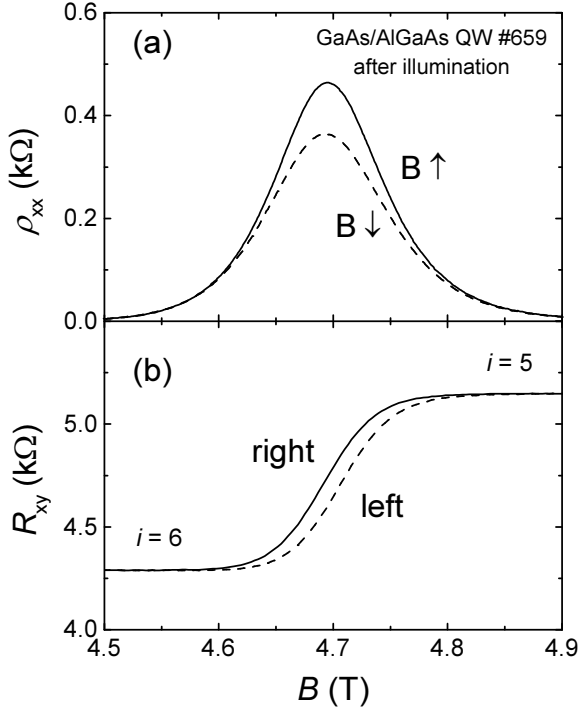}
\caption{\label{fig:Long GaAs after} Longitudinal resistivity and
Hall resistance of the GaAs/AlGaAs quantum well (\#659) after
illumination ($T = 0.4$ K). Data are shown for the $i = 6
\rightarrow 5$ plateau-plateau transition. Upper panel: $\rxx$
data measured at the same contact pair for two directions of $B$
as indicated. Lower panel: $R_{xy}$ data measured at right and
left Hall contact pairs as indicated.}
\end{figure}
After illumination, the GaAs/AlGaAs quantum well becomes more
homogenous as expected. Magnetotransport data near the $i =
6\rightarrow 5$ PP transition are shown in Fig.~\ref{fig:Long GaAs
after}. The carrier density increases from 4.7 to $6.1 \times
10^{11}$ cm$^{-2}$, while the carrier difference decreases to
$\Delta n_e / n_e = 0.0025$. The longitudinal resistance still
remains antisymmetric, but the effect is now much smaller and
amounts to only 20\% for the $i = 6 \rightarrow 5$ PP transition.

\begin{figure}[htb]
\includegraphics[width=7.6cm]{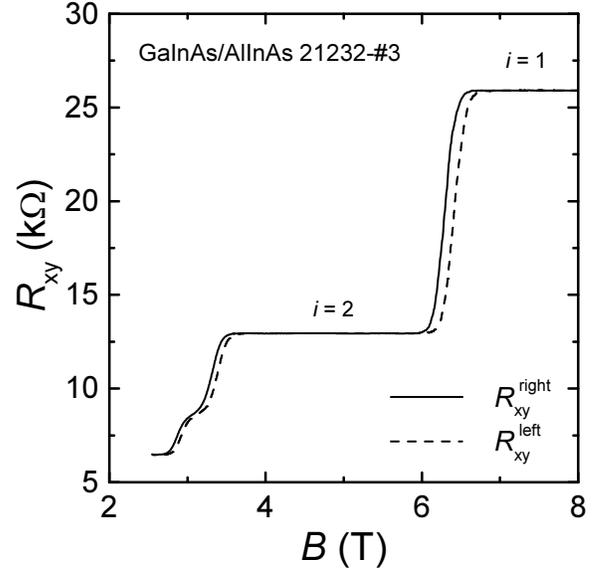}
\caption{\label{fig:Hall GaInAs} Hall resistance as a function of
magnetic field for the GaInAs/AlInAs quantum well (21232-\#3 - no
illumination) near plateaux $i = 1-3$ at $T = 0.4$ K. The solid
and dashed lines show data taken at the right and left Hall
contact pairs.}
\end{figure}

\begin{figure}[htb]
\includegraphics[width=7.6cm]{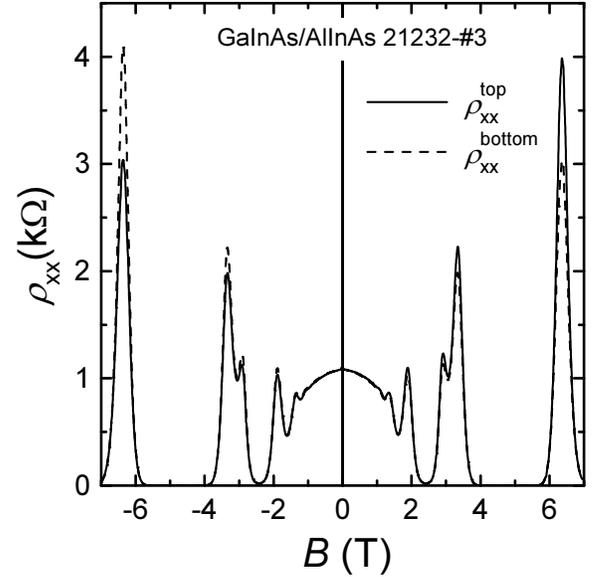}
\caption{\label{fig:Long GaInAs} Longitudinal resistivity as a
function of magnetic field ($B\uparrow$ and $B\downarrow$) for the
GaInAs/AlInAs quantum well (21232-\#3 - no illumination) at $T =
0.4$ K. The solid and dashed lines show data taken at contact
pairs located at the top and bottom side of the Hall bar.}
\end{figure}

Magnetotransport data for the GaInAs/AlInAs quantum well are shown
in Figs.~\ref{fig:Hall GaInAs} and~\ref{fig:Long GaInAs}.
Again we show the data before illumination ($n_e = 1.8 \times
10^{11}$ cm$^{-2}$). From Fig.~\ref{fig:Hall GaInAs} we extract
$\Delta n_e / n_e = 0.014$. The antisymmetry effect
(Fig.~\ref{fig:Long GaInAs}) is not as pronounced as for the
GaAs/AlGaAs quantum well, but nevertheless significant. After
illumination ($n_e = 3.6 \times 10^{11}$ cm$^{-2}$) $\Delta n_e /
n_e = 0.006$ (Table~\ref{table}) and the antisymmetry effect is
reduced. Magnetotransport measurements on a second Hall bar,
prepared from the same wafer, with $n_e = 2.2 \times 10^{11}$
cm$^{-2}$ show a comparable $\Delta n_e / n_e \sim 0.018$ (see
Table~\ref{table}) and the corresponding antisymmetry effect.

Finally, we mention, that we have also investigated an InGaAs/InP
heterostructure. This sample with $n_e = 2.2 \times 10^{11}$
cm$^{-2}$ and $\mu  = 16000$ cm$^2 /$Vs has been used previously
for magnetotransport studies of the plateau-plateau and
plateau-insulator transitions in the quantum Hall regime
(Refs.~\onlinecite{hwang:prb93, schaijk:prl00, lang:pe02,
stretch}). The Hall data show that it has a fairly large carrier
density difference $\Delta n_e / n_e \sim 0.016$. Although
$R_{xx}^b$ and $R_{xx}^t$ differ considerably the data are not
antisymmetric, thus $R_{xx}^b(B) \neq R_{xx}^t(-B)$. This is the
result of an additional ``misalignment'' of the voltage contacts
due to intrinsic inhomogeneities (i.e. different effective values
of $L$).

\section{The effect of a carrier density gradient}

In the following we show that the observed asymmetry effect can be
accounted for by a small carrier density gradient in the sample.
Let us assume that the carrier density gradient points from the
left current contact to the right one. In this case, $R_{xy}^l$
and $R_{xy}^r$ are slightly different. We define
\begin{equation}
\Delta R_{xy} = R_{xy}^l - R_{xy}^r,
\end{equation}
and for the difference in the longitudinal resistances measured at
both sides of the Hall bar
\begin{equation}
\Delta R_{xx} = R_{xx}^t - R_{xx}^b.
\end{equation}
For a perfect Hall bar $\Delta R_{xy} = 0$ and $\Delta R_{xx} =
0$. Since
\begin{eqnarray}
\Delta R_{xy} = ((V_3 - V_5) - (V_4 - V_6)) / I = \nonumber
\\ ((V_3 - V_4) - (V_5 - V_6)) / I = \Delta R_{xx},
\label{TwoDeltas}
\end{eqnarray}
a finite $\Delta R_{xy}$ immediately implies that there is a
difference between the longitudinal resistances measured at the
top and bottom side of the Hall bar, $\Delta R_{xx} \neq 0$.
Because the Hall voltage is an odd function of the magnetic field,
reversing the polarity changes the sign of both $\Delta R_{xy}$
and $\Delta R_{xx}$:
\begin{equation}
\Delta R_{xy} (- B) = - \Delta R_{xy} (B),
\end{equation}
and
\begin{equation}
\Delta R_{xx} (- B) = - \Delta R_{xx} (B).
\end{equation}
Thus $\Delta R_{xx}$ is an antisymmetric function, which holds
under the conditions of eq.~(\ref{antisymmetry}).
Eq.~(\ref{TwoDeltas}) may be directly verified by comparing the
differences $\Delta R_{xx} (B)$ and $\Delta R_{xy} (B)$ calculated
from the measured data (compare for instance $\Delta R_{xx} =
(L/W) \Delta \rho_{xx}$ in Fig.~\ref{fig:Long GaAs after}a with
$\Delta R_{xy}$ in Fig.~\ref{fig:Long GaAs after}b).

\section{Plateau-plateau transitions}

The effect of a carrier density gradient on the magnetotransport
data as described here is in fact generally applicable to any 2D
and 3D system, as follows from Eq.(\ref{TwoDeltas}). However, the
effect can be dramatically large for 2D quantum Hall samples,
because of the steep slope of $R_{xy}(B)$ at the plateau-plateau
(PP) transitions. Consequently, our findings yield important
constraints on the construction of the conductivity flow diagrams
of the PP transitions~\cite{pruisken:prl88}. In the following
paragraphs, a formal treatment of magnetotransport at the PP
transition in the presence of a small carrier density gradient is
presented.

We start from the usual equations for the Hall bar geometry, which
tell us that a uniform current density $j_x$ results from an
applied electric field $E_x$ along the channel direction ($x$):
\begin{eqnarray}
E_x = \rho_0 j _x, &E_y = \rho_H j_x.
\end{eqnarray}
Here $j_y = 0$ and $\rho_0$ and $\rho_H$ are the longitudinal and
Hall resistivity for a perfectly homogeneous sample. In a simple
model for sample inhomogeneities we assume that the critical field
$B^*$ or filling fraction $\nu^*$ for the PP transition varies
linearly in $x$ due to a gradient in the carrier density. At the
PP transition the Hall resistivity slopes from one Hall plateau to
the other, while the longitudinal resistivity forms a peak, so
that $|\partial \rho_H / \partial B| \gg |\partial \rho_0 /
\partial B|$. As such, an $x$-dependence of the local filling factor
$\nu(x)$ therefore mainly affects the electric field component
$E_y$:
\begin{eqnarray}
E_x = \rho_0 j_x,& E_y = (\rho_H + \alpha x ) j_x, &\alpha =
\frac{\partial \rho_H}{\partial \nu} \frac{\partial \nu}{\partial
x}.
\end{eqnarray}
This result, however, violates an important condition for having a
stationary state, i.e. the electric field must be rotation free
$\mathbf{\nabla \times E} = 0$. To satisfy this condition we
proceed by inserting a $y$-dependent current density $j_x
\rightarrow j_x ( 1 + \beta y)$, where $\beta = \alpha / \rho_0$.
Working to linear order in the coordinates $x$ and $y$ we can
write
\begin{eqnarray}
E_x = (\rho_0 + \rho_0 \beta y) j_x, &E_y = (\rho_H + \alpha x +
\rho_H \beta y) j_x.
\end{eqnarray}
Hence
\begin{eqnarray}
E_x = (\rho_0 + \alpha y) j_x, &E_y = (\rho_H + \alpha x + \alpha
\frac{ \rho_H}{\rho_0} y) j_x,
\end{eqnarray}
are the appropriate equations for the PP transition. Notice that
the stationary state condition $\mathbf{\nabla \times E} = 0$  and
charge conservation $\nabla \mathbf{\cdot j} = 0$  are satisfied.

The longitudinal resistance at the top and bottom and the Hall
resistance at the left and right contacts of the Hall bar are
given by
\begin{subequations}
\label{Rxx}
\begin{equation}
R_{xx}^t = V_{xx}^t / I_x = \frac{L}{W}(\rho_0 + \alpha
\frac{W}{2}),
\end{equation}
\begin{equation}
R_{xx}^b = V_{xx}^b / I_x = \frac{L}{W}(\rho_0 - \alpha
\frac{W}{2}),
\end{equation}
\end{subequations}
\begin{subequations}
\label{rho_xy}
\begin{equation}
R_{xy}^l = V_{xy}^l / I_x = (\rho_H - \alpha \frac{L}{2}),
\end{equation}
\begin{equation}
 R_{xy}^r = V_{xy}^r / I_x = (\rho_H + \alpha
\frac{L}{2}), %
\end{equation}
\end{subequations}
where we take zero coordinates $(x, y)$ in the center of the Hall
bar. When $B \rightarrow - B$, $\rho_H$ and $\alpha$ change sign,
but $\rho_0$ remains unchanged. The results therefore explain the
observed asymmetry at the PP transition, $R_{xx}^b(B^*) =
R_{xx}^t(-B^*)$. Notice that with this specific form of $\alpha$,
eq.(\ref{rho_xy}) can be regarded as a Taylor expansion of a local
Hall resistance $\rho_H$ at values $x = \pm L/2$.

\section{Discussion}
\begin{figure}[htb]
\includegraphics[width=7.6cm]{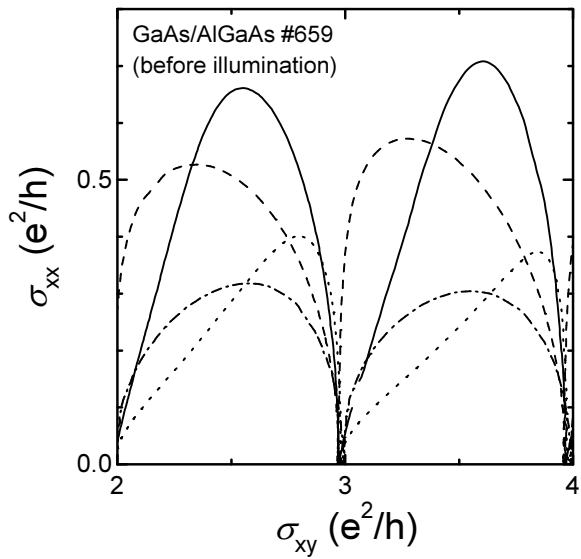}
\caption{\label{fig:nonideal} The longitudinal conductance $\sxx$
as a function of the Hall conductance $\sxy$ for the $i = 4
\rightarrow 3$ and $3 \rightarrow 2$ PP transitions of the
GaAs/AlGaAs quantum well before illumination ($T = 0.4$ K). The
four different lines are obtained by using all possible
combinations of $\rxx$ and $R_{xy}$.}
\end{figure}

\begin{figure}[htb]
\includegraphics[width=7.6cm]{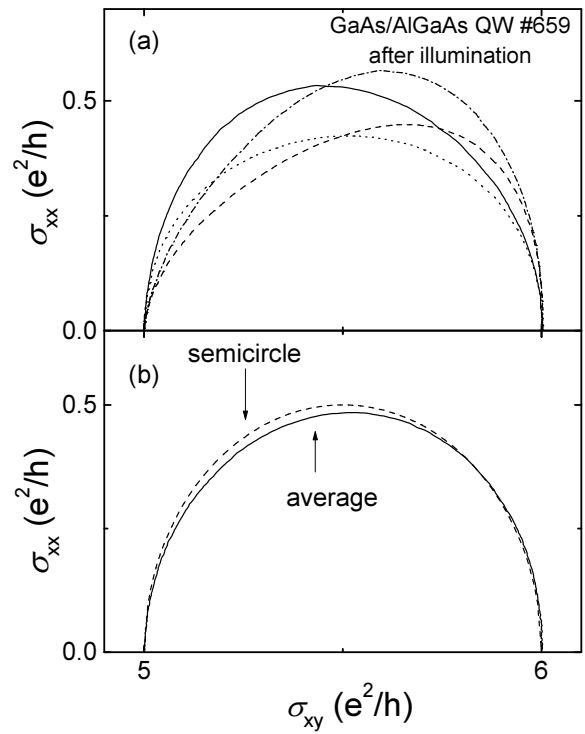}
\caption{\label{fig:ideal} (a) The longitudinal conductance $\sxx$
as a function of the Hall conductance $\sxy$ for the $i = 5
\rightarrow 6$ PP transition of the GaAs/AlGaAs quantum well after
illumination ($T = 0.4$ K). The four different lines are obtained
by using all possible combinations of $\rxx$ and $R_{xy}$. (b) The
solid line represents the $\sxx,\sxy$ data after averaging. The
dashed line shows the ``ideal'' semicircle relation. }
\end{figure}

Eqs.(\ref{Rxx})-(\ref{rho_xy}) show that the probed Hall
resistance $R_{xy}$ is a local resistance, and that the
longitudinal resistance $R_{xx}$ is not identical under field
reversal. This gives rise to complications in extracting the
conductivity components $\sigma_{xx} = \rho_{xx}/(\rho_{xx}^2 +
\rho_{xy}^2)$ and $\sigma_{xy} = \rho_{xy}/(\rho_{xx}^2 +
\rho_{xy}^2)$ of the PP transitions. As an example we show in
Fig.~\ref{fig:nonideal} the $\sigma_{xx}$,$\sigma_{xy}$
conductance plane with data for the $i = 4 \rightarrow 3$ and $3
\rightarrow 2$ PP transitions of the GaAs/AlGaAs quantum well
(before illumination).
By choosing different combinations of $R_{xx}^t$ or $R_{xx}^b$ and
$R_{xy}^l$ or $R_{xy}^r$ we obtain four different $\sigma_{xx}$,
$\sigma_{xy}$ curves for one field direction (for the reverse
field direction identical curves are obtained, but the $R_{xx}$, $
R_{xy}$ labelling is different). Clearly, all curves show strong
deviations from the ``ideal'' semicircle
relation~\cite{pruisken:prl88,dykhne:prb94} $\sigma_{xx}^2
+(\sigma_{xy} - n\sigma^*)^2 = (\sigma^*)^2$
 with $\sigma^* = e^2 / 2h$ and
$n =$ 7 and 5 for the $i = 4 \rightarrow 3$ and $3 \rightarrow 2$
PP transitions, respectively. After illumination, the density
increases and the sample becomes more homogeneous as can be
concluded from Fig.~\ref{fig:ideal}, where we show $\sigma_{xx},
\sigma_{xy}$ for the $i = 6 \rightarrow 5$ PP transition.
After averaging the different curves, $\sigma_{xx}, \sigma_{xy}$
follows quite close the semicircle relation with $n = 11$.

Temperature driven $\sxx, \sxy$ flow diagrams are an important
experimental test of the two-parameter renormalization group
theory of the quantum Hall effect~\cite{pruisken:prl88,wei:prb85}.
In the scaling theory of the quantum Hall
effect~\cite{pruisken:prl88} the critical resistivity
$\rho_0(B^*)$ (or critical conductivity $\sigma_0(B^*) =
\sigma^*$) at the PP transition is a constant when $T \rightarrow
0$. The presence of a carrier density gradient, therefore, brings
about a second complication in analysing magnetotransport data, as
the parameter $\alpha$ varies with temperature through $\partial
\rho_H / \partial \nu$, namely a strong temperature variation of
$R_{xx}^{t,b}(B^*)$. The temperature dependence of the measured
critical resistance at the PP transition, eq.~(\ref{Rxx}), can be
written as
\begin{equation}
R_{xx}^{t, b}(B^*, T) \propto (1 \pm \mathit{const} \cdot
\alpha(T))
\end{equation}
Notice that when $B \rightarrow -B$ the amplitudes change
according to $R_{xx}^t (B^*, T) =  R_{xx}^b(-B^*, T)$.
Corresponding changes in amplitude occur in the conductivities
$\sxx (B^*)$ evaluated from $R_{xx}^t$ or $R_{xx}^b$. This feature
of density gradients has been noticed in InGaAs/InP
heterostuctures~\cite{lang:tba}. The strong temperature variation
of $\sxx (B^*)$ is removed at low temperatures when averaging the
data sets obtained for both directions of the magnetic field. Thus
reversing the magnetic field is of paramount importance in
determining the proper temperature variation of the conductivity
peak height $\sxx (B^*)$. The temperature variation of $\sxx
(B^*)$, reported in the literature at several
places~\cite{wei:prb92,rokhinson:ssc95}, might thus well be the
result of not averaging the magnetotransport data over up and down
magnetic field directions. This suggests that the occurrence of a
``non-universal'' critical conductivity can be explained without
the need to evoke percolation models with macroscopic fluctuations
of the local filling factor~\cite{ruzin:prb96}.

At this point it is important to stress that the experimental data
reported in Figs.~\ref{fig:nonideal} and~\ref{fig:ideal} are taken
on a GaAs/AlGaAs quantum well. In general, in GaAs based quantum
wells and heterostructures the dominant scattering mechanism is
long-ranged due to remote impurity doping. This complicates the
observation of universal scaling laws in the quantum Hall effect,
as the scaling behavior is pushed to very low temperatures, where
it is difficult to access experimentally~\cite{pruisken:prb99}. At
the ``elevated'' temperatures used here thermal smearing due to
the Fermi-Dirac distribution results in transport properties
semi-classical in nature. The observation of a near ``ideal''
semicircle relation (Fig.~\ref{fig:ideal}) in our GaAs/AlGaAs
quantum well is therefore not a signature of scaling.

The analysis of the PP transition has been made for Hall bars with
a constant density gradient along the channel direction. In a real
sample the situation may be more complex (gradients in $x$-$y$
directions, non-linear density variations). In addition, intrinsic
sample inhomogeneities can also lead to Hall potential contact
misalignment, which in general gives rise to a contribution from
$R_{xx}$ to the $R_{xy}$ data. In Ref.~\cite{stretch} a novel
analytical procedure was presented to disentangle the universal
quantum critical aspects of the magnetotransport data and sample
dependent aspects, such as density gradients and contact
misalignment. The methodology is based on decomposing the measured
$\rho_{ij}$ in symmetric and antisymmetric parts, as made here for
the most simple case. So far, the analysis has been applied
successfully to the plateau-insulator (PI) phase transition
measured for an InGaAs/InP
heterostructure~\cite{stretch,lang:tba}. At the PI transition,
$R_{xy} \sim h/e^2$ remains quantized which facilitates the
analysis. At the PP transition the situation is much more complex,
as both $\rxx$ and $\rxy$ are a function of $T$ and $B$. In this
case, higher order contributions to $n_e(x)$ have to be considered
as well.

\section{Conclusions}
The effect of macroscopic sample inhomogeneities on
magnetotransport data of the two-dimensional electron gas in the
Hall bar geometry has been investigated. We find a remarkable
antisymmetry in the longitudinal resistances $R_{xx}^t$ and
$R_{xx}^b$ measured on both sides of the Hall bar upon reversing
the magnetic field: $R_{xx}^t (B) = R_{xx}^b (-B)$. The
antisymmetry in $R_{xx}$ is explained by a small carrier gradient
along the channel direction of the Hall bar. The presence of a
carrier density gradient complicates the study of quantum
criticality of the plateau-plateau transitions in the quantum Hall
effect. We evaluate expressions for the resistances at the
plateau-plateau transitions and demonstrate complications that
arise in extracting $\sxx, \sxy$ flow diagrams.

Finally, we mention that in order to obtain an experimentally
accessible large temperature range for scaling, the dominant
electron scattering mechanism should be provided by short-range
potential fluctuations~\cite{pruisken:prb99}, as for instance is
the case for alloy scattering. From the point of view of sample
preparation, this is virtually inherent to macroscopic sample
inhomogeneities. Any proper study of quantum criticality of the
quantum Hall effect therefore requires an unremitting research
effort in understanding and modelling the effect of macroscopic
sample inhomogeneities.

\begin{acknowledgments}
This work was supported by FOM (Dutch Foundation for Fundamental
Research of Matter). V.A.K. and  G.B.G were supported by research
grant RFBR 00-02-17493. The authors would like to thank Peter
Nouwens from the Technical University Eindhoven for the
preparation of Hall bars and Paul Koenraad for valuable
discussions.
\end{acknowledgments}



\end{document}